\documentclass[aps,prl,twocolumn]{revtex4-1}

\usepackage{fancyhdr}
\usepackage{amsmath}
\usepackage{graphicx}
\usepackage{lipsum}

\newcommand{\ket}[1]{\ensuremath{\left|#1\right\rangle}}

\newcommand{\reverse}{\ensuremath{(0,2) \rightarrow (1,1)~}}
\newcommand{\forward}{\ensuremath{(1,1) \rightarrow (0,2)~}}
\newcommand{\charge}{\ensuremath{(1,1) \leftrightarrow (0,2)~}}
\newcommand{\muspace}{\ensuremath{\mu}-space~}

\begin{document}

\title{A new Regime of Pauli-Spin Blockade}
\author{Justin K. Perron}
\email{Perronjk@gmail.com}
\affiliation{California State University San Marcos, Department of Physics, California 92096}
\affiliation{Joint Quantum Institute, University of Maryland}
\affiliation{National Institute of Standards and Technology}
\author{M. D. Stewart, Jr.}
\author{Neil M. Zimmerman}
\affiliation{National Institute of Standards and Technology}
\date{\today}

\begin{abstract}
	Pauli-spin blockade (PSB) is a transport phenomenon in double quantum dots that allows for a type of spin to charge conversion often used to probe fundamental physics such as spin relaxation and singlet-triplet coupling. In this paper we theoretically explore Pauli-spin blockade as a function of magnetic field $B$ applied parallel to the substrate. In the well-studied low magnetic field regime, where PSB occurs in the forward \forward tunneling direction, we highlight some aspects of PSB that are not discussed in detail in existing literature, including the change in size of both bias triangles measured in the forward and reverse biasing directions as a function of $B$. At higher fields we predict a crossover to ``reverse PSB'' in which current is blockaded in the reverse direction due to the occupation of a spin singlet as opposed to the traditional triplet blockade that occurs at low fields.  The onset of reverse PSB coincides with the development of a tail like feature in the measured bias triangles and occurs when the Zeeman energy of the polarized triplet equals the exchange energy in the (0,2) charge configuration.  In Si quantum dots these fields are experimentally accessible; thus, this work suggests a way to probe singlet to triplet relaxation mechanisms in quantum dots when both electrons occupy the same quantum dot.
	 
\end{abstract}

\maketitle


Since its discovery, Pauli-spin blockade (PSB)\cite{Ono02} has been a valuable tool for probing fundamental physics.  Its use in spin to charge conversion has led to investigations of spin $T_1$ relaxation times\cite{Petta05B,Hu12}, electron spin couplings to lattice nuclear spins\cite{Johnson05A}, and spin-orbit effects\cite{Nadj10}.  PSB has also received a lot of attention from the quantum information community due to its use in read-out and initialization of various electron spin states\cite{Petta05,Koppens06,Nowack07}.  The majority of studies involving PSB have focused on the low magnetic field regime. In this manuscript we investigate the qualitative behavior of PSB at all magnetic fields. In doing so we identify experimentally accessible regimes where PSB has yet to be studied and new physics is likely to be found.

In a typical double quantum dot (DQD) device two quantum dots are coupled to each other as well as to two electron reservoirs (see figure~1(a)). The dots are also capacitively coupled to one or more gate electrodes which can be used to control the dot potentials. The charge configuration of the system is characterized by a pair of numbers $(n,m)$ corresponding to the number of electrons on the left and right dot, respectively. Transport through quantum dots can be understood by considering the chemical potentials of the quantum dots, $\mu_{l}(n,m) \equiv E(n,m) - E(n-1,m)$ and $\mu_{r}(n,m) \equiv E(n,m)-E(n,m-1)$, where $E$ denotes the total energy of the system. When the appropriate chemical potentials form a energetically downhill path between the Fermi levels of the two leads, $\mu_{LL(RL)}$, electron tunnelling through the device is energetically favorable and leads to current flow\cite{VDW02}. Near the \charge charge transition, where PSB typically occurs, this condition can be expressed as $\mu_{LL} \geq \mu_l(1,1) \geq\mu_r(0,2) \geq \mu_{RL}$ for transport in the forward \forward direction and $\mu_{LL} \leq \mu_l(1,1) \leq\mu_r(0,2) \leq \mu_{RL}$ for transport in the reverse \reverse direction.  These conditions describe regions of transport known as bias triangles, with sizes proportional to the applied bias voltage, $V_{bias}$.\cite{VDW02}

For a more complete picture of transport through a DQD near the \charge charge degeneracy we now also consider the spin state of the electrons.  For transport in the forward (left to right) direction current flows via tunneling events that cycle through the $(0,1) \rightarrow (1,1) \rightarrow (0,2) \rightarrow (0,1)$ charge configurations. In the reverse direction it occurs via $(0,1) \rightarrow (0,2) \rightarrow (1,1) \rightarrow (0,1)$.  In both cases the first tunneling event brings an electron from a lead onto the DQD forming a two-electron system.  These electrons can form one of four possible spin states, the $s=0$ spin singlet $\left|S\right\rangle~\equiv~\left(\left|\uparrow\downarrow\right\rangle - \left|\downarrow\uparrow\right\rangle\right)/\sqrt{2}$, or one of the three $s=1$ triplet states $\left|T_0\right\rangle~\equiv ~\left(\left|\uparrow\downarrow\right\rangle + \left|\downarrow\uparrow\right\rangle\right)/\sqrt{2}$, $\left|T_+\right\rangle~\equiv~\left|\uparrow\uparrow\right\rangle$, $\left|T_-\right\rangle~\equiv~\left|\downarrow\downarrow\right\rangle$. In zero magnetic field, the \ket{S} state is typically the ground state and the three degenerate triplet states are higher in energy by the exchange energy $J$\cite{AshcroftMermin}. Since $J$ scales with the wavefunction overlap between the two electrons, the charge configuration of the electrons will have a large influence on the magnitude of $J$. When in the (0,2) charge configuration, the overlap is large so $J(0,2)$ is as well, whereas in the (1,1) configuration the overlap is minimal as is $J(1,1)$\footnote{In this paper we take $J(1,1) = 0$ for simplicity. Including a finite value would not significantly affect the results of this work.}. This difference in exchange between the two charge configurations is the cornerstone of PSB.  It leads to a situation where although the forward \forward interdot transition is allowed for the \ket{S} ground states, current does not flow.  This occurs because, with $J(1,1) < J(0,2)$, it is possible that even though conduction through the singlet states is allowed, {\emph{i.e.}} $\mu_{LL}\geq\mu_l^{\left|S\right\rangle}(1,1) \geq \mu_r^{\left|S\right\rangle}(0,2)\geq \mu_{RL}$, the interdot transition for the triplet states is forbidden, $\mu_l^{\left|T_x\right\rangle}(1,1)\leq\mu_r^{\left|T_x\right\rangle}(0,2)$.  Thus, once loaded, the \ket{(1,1)T_x} can tunnel neither forward nor back until it has relaxed into a singlet state. Since the spin relaxation from a \ket{T_x} state to a \ket{S} state is typically long relative to the tunneling times, the measured current in the PSB region is near or below the noise floor of the measurement system. This situation truncates the size of the bias triangles in the forward bias direction, as shown in figure~1c, while leaving the triangles measured in the reverse bias direction at ``full size'', as in figure~1b. This size difference is the most basic signature of PSB and has been measured by several groups in various material systems\cite{Johnson05B,Lai11,Yamahata12,Simmons10,Koppens05,Shaji08,Churchill09A,Weber14,Liu08,Pfund07,Nadj10}.

\begin{figure}
	\centering
	\includegraphics[width = 0.9\columnwidth]{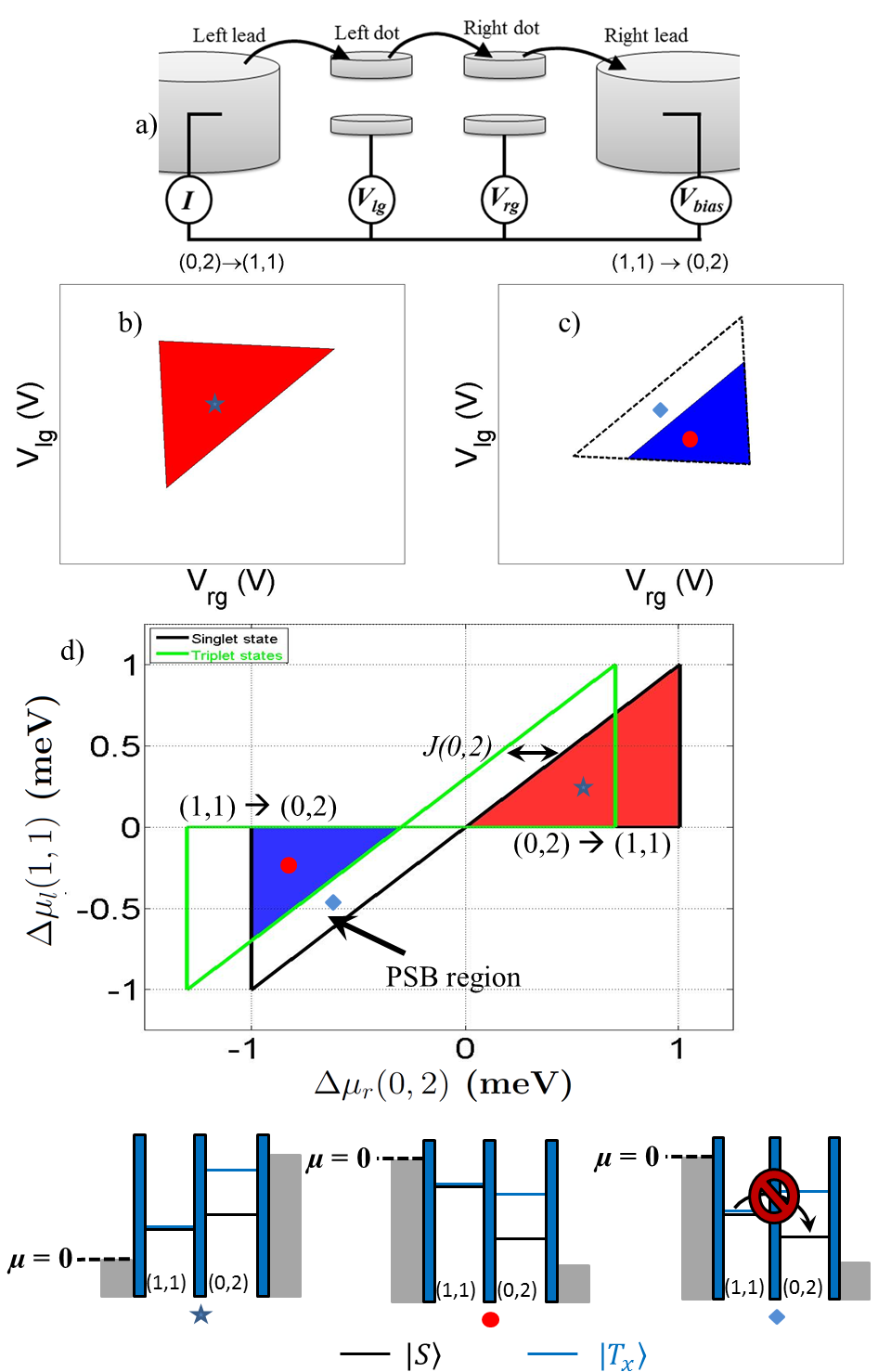}   
	\caption{{\bf{Traditional forward Pauli-spin blockade, B = 0 T}} Bias triangles at the $(1,1) \leftrightarrow (0,2)$ charge transition with no applied magnetic field, $V_{bias} = \pm 1~{\rm{mV}}$ and $J(0,2) = 0.3~{\rm{meV}}$. (a) Schematic of a typical double quantum dot. A bias applied between the left and right leads can lead to electrons tunneling through the device. The potentials of the dots are controlled with two capacitively coupled gates. (b) and (c) are voltage space pictures of current measurements in the \reverse and \forward biasing directions using typical voltage and capacitance values. Filled regions of color correspond to non-zero current. (d) is the chemical potential space picture for the same transitions. Triangles below the $\Delta\mu_{l}(1,1) = 0$ axis are for \forward biasing and those above are for \reverse biasing. Chemical potential diagrams for certain regions of interest in (b), (c) and (d) are also shown. The \muspace origin is determined by the Fermi level of the left lead which we take to be at the ground potential for all measurements.  The PSB region is labelled with the blue diamond.\label{fig1}}
\end{figure}

For a more intuitive presentation of the voltage space measurements depicted in figure~1(b) and (c) one can convert them to chemical potential space, or \muspace\cite{Perron15}, where triangles from both bias polarities can be shown on a single plot. Plotted in figure~1(d) is the \muspace equivalent to what is shown in both panels~(b) and (c), where the axes are defined as $\Delta\mu_l(1,1) \equiv \mu_l^{\left|S\right\rangle}(1,1) - \mu_{LL}$ and $\Delta\mu_r(0,2) \equiv \mu_r^{\left|S\right\rangle}(0,2) -\mu_{LL}$.  Also shown in this plot are outlines of the ``state triangles'', which correspond to the conditions $\mu_{LL} \geq \mu_{l}^{state}(1,1) \geq \mu_{r}^{state}(0,2) \geq \mu_{RL}$, in the forward biased direction and $\mu_{RL} \geq \mu_{r}^{state}(0,2) \geq \mu_{l}^{state}(1,1) \geq \mu_{LL}$, with $\mu_{LL} \equiv 0$,  in the reverse direction, where $state$ refers to the spin state, \ket{S} or \ket{T_x}, of the two electron system. In figure~1(c) and (d), the region labeled with the blue diamond corresponds to the region where PSB occurs, as shown in the corresponding chemical potential diagram.

When a magnetic field $B$, is applied, several effects can occur depending on the direction of the applied field. If applied perpendicular to the two-dimensional electron gas (2DEG) there can be a significant change in the spatial wavefunctions of the electrons. This will in turn affect the magnitudes of $J(1,1)$ and $J(0,2)$ leading to a change in size of the PSB region\cite{Johnson05B}. At higher fields, the \ket{T_x} states can fall below the \ket{S} states leading to PSB in the \forward direction caused by singlet states rather than the triplet states\cite{Sun12}. When $B$ is applied in the plane of the 2DEG the spatial effect on the wave functions is negligible and observed effects are due to the Zeeman splitting of the triplet levels. This in-plane magnetic field dependence will be the focus of the remainder of this manuscript.

When considering the in-plane $B$, we make the following assumptions for $B$-dependence of the chemical potentials
\begin{align}
	\left. \mu^{\left|S\right\rangle}\left(n,m\right)\right|_{B\neq0} &= \left.\mu^{\left|S\right\rangle}\left(n,m\right)\right|_{B=0}, \\
	\left. \mu^{\left|T_0\right\rangle}\left(n,m\right)\right|_{B\neq0} &=\left. \mu^{\left|S\right\rangle}\left(n,m\right)\right|_{B=0} + J(n,m), \\
	\left.	\mu^{\left|T_\pm\right\rangle}\left(n,m\right)\right|_{B\neq0} &=\left. \mu^{\left|S\right\rangle}\left(n,m\right)\right|_{B=0} + J(n,m) \pm g\mu_BB.
\end{align}
\noindent When $B$ is increased from zero, the first notable effect is the change in size of the bias triangles. As reported in reference~\cite{Lai11}, the triangles in the forward \forward biasing direction grow in size proportional to the Zeeman energy $E_Z = g\mu_B B$, with $g$ the electron g-factor, and $\mu_B$ the Bohr magneton.  However, we predict a concomitant reduction in size of the triangles measured in the reverse \reverse direction.  Both of these effects are illustrated in figure~2. 

Figure~2(a) shows the \muspace picture of the changing triangles (see supplemental material for details on using the \muspace picture to determine the regions of allowed current).  The origin is defined by the point where $\mu_l^{\left|S\right\rangle}(1,1) =\mu_r^{\left|S\right\rangle}(0,2) = \mu_{LL} = 0$. Since $\mu^{\left|S\right\rangle}(n,m)$ has no magnetic field dependence, our definition of the origin relates it to a single point in voltage space, at all magnetic fields. This has the advantage that the direction the growth/contraction occurs along is clarified. The triangles for both biasing directions change size along the $\Delta\mu_{l}(1,1)$ axis. This is understood by considering the process which determines the edge of the triangles in each case.  For the forward \forward direction, the crucial process is the loading of an electron from the left lead onto the left dot.  As $B$ is increased, the \ket{(1,1)T_-} level is lowered by the Zeeman energy meaning that the loading of that state can occur at larger $\Delta\mu_l(1,1)$ values. This is shown in the chemical potential diagrams at the top of figure~2a for the point labelled with the blue triangle. In this region, since the only accessible (1,1) state is \ket{T_-}, conduction is spin-polarized. In the reverse \reverse direction, the crucial process is the Coulomb blockade of current due to loading of the \ket{(1,1)T_-} level below $\mu_{LL}$. Again, as $B$ is increased this level drops below $\mu_{LL}$ at a higher $\Delta\mu_l(1,1)$. This is depicted in the chemical potential diagrams at the bottom of figure~2a for the point labeled with the red square. In both biasing directions, when viewed in this \muspace picture, the direction the triangles are changing is clear, and we can then easily generate the correct gate voltage space picture. This is done in figure~2(b)~ through~(d), using typical capacitance and voltage parameters, which show bias triangles for the forward \forward direction growing with magnetic field in roughly the $-V_{lg}$ direction (as compared to the detuning direction i.e. perpendicular to the base of the triangles). This prediction is clearly observed in reference~\cite{Lai11} and validates the assumptions made in equations~1 through 3. 


\begin{figure}
	\centering
	\includegraphics[width = 0.9\columnwidth]{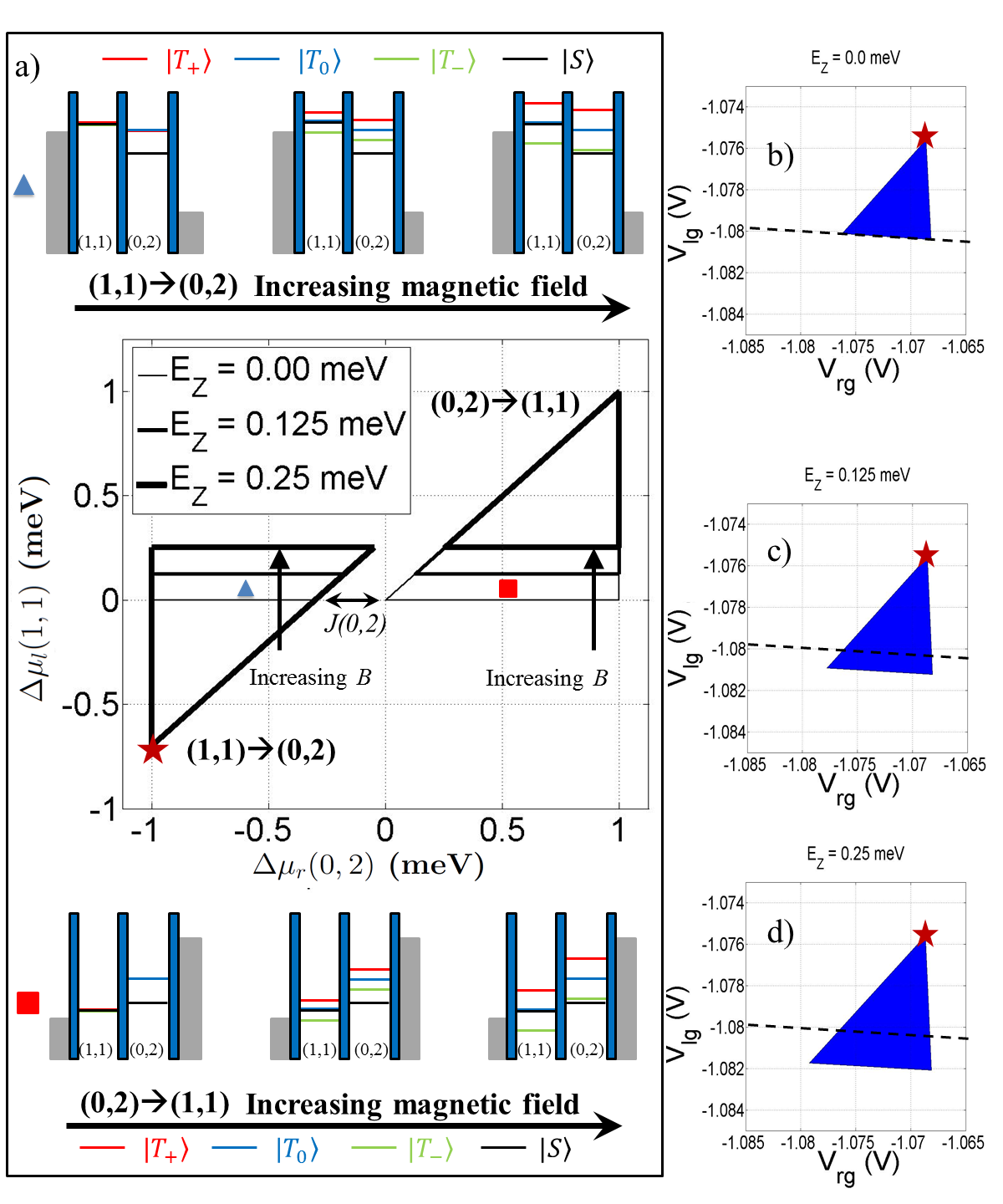}  
	\caption{{\bf{Magnetic field dependence for $\mathbf{B \leq B_{ST}}$}}: (a) \muspace picture of bias triangles in both biasing directions with $V_{bias}~=~\pm1~\rm{mV}$, $J(0,2) = 0.3$~meV at three $B$ values corresponding to $E_Z = 0.0$, 0.125 and 0.25~meV respectively. The triangles outlined with consecutively thicker lines correspond to higher $B$. As $B$ is increased, the triangles in the forward \forward direction (left set of triangles) grow along the $\mu_l(1,1)$ direction while the reverse \reverse biased triangles shrink in the same direction.  The change in triangle size is due to the Zeeman splitting of the \ket{(1,1)T_-} level as shown in the chemical potential diagrams corresponding to the points labeled with the blue triangle and the red square respectively.  (b) through (d) are gate voltage space bias triangles in the forward \forward biasing direction at the same magnetic fields as (a). The vertex labeled with the star occurs at the same position at all magnetic fields in both \muspace and gate voltage space. The growth direction we predict is clear and agrees with the data in Figure~4 of reference~\cite{Lai11}.\label{fig1}}
\end{figure}

As $B$ is increased further there will be a crossover from PSB in the forward \forward direction, as discussed above and in previous works\cite{Ono02,Nadj10,Liu08,Lai11,Yamahata12,Johnson05B,Koppens05,Simmons10,Shaji08,Churchill09A,Churchill09B,Weber14,Liu08,Pfund07,Sun12}, to PSB in the reverse \reverse direction.  The onset of this reverse PSB is accompanied by the development of a tail-like feature in the regions of allowed current\footnote{These tail features are qualitatively similar to those associated with lifetime enhanced transport\cite{Shaji08,Simmons10} but arise from very different physical mechanisms.} and occurs at a magnetic field $B_{ST}$ where $E_Z = J(0,2)$. 

\begin{figure}
	\centering
	\includegraphics[width = 0.9\columnwidth]{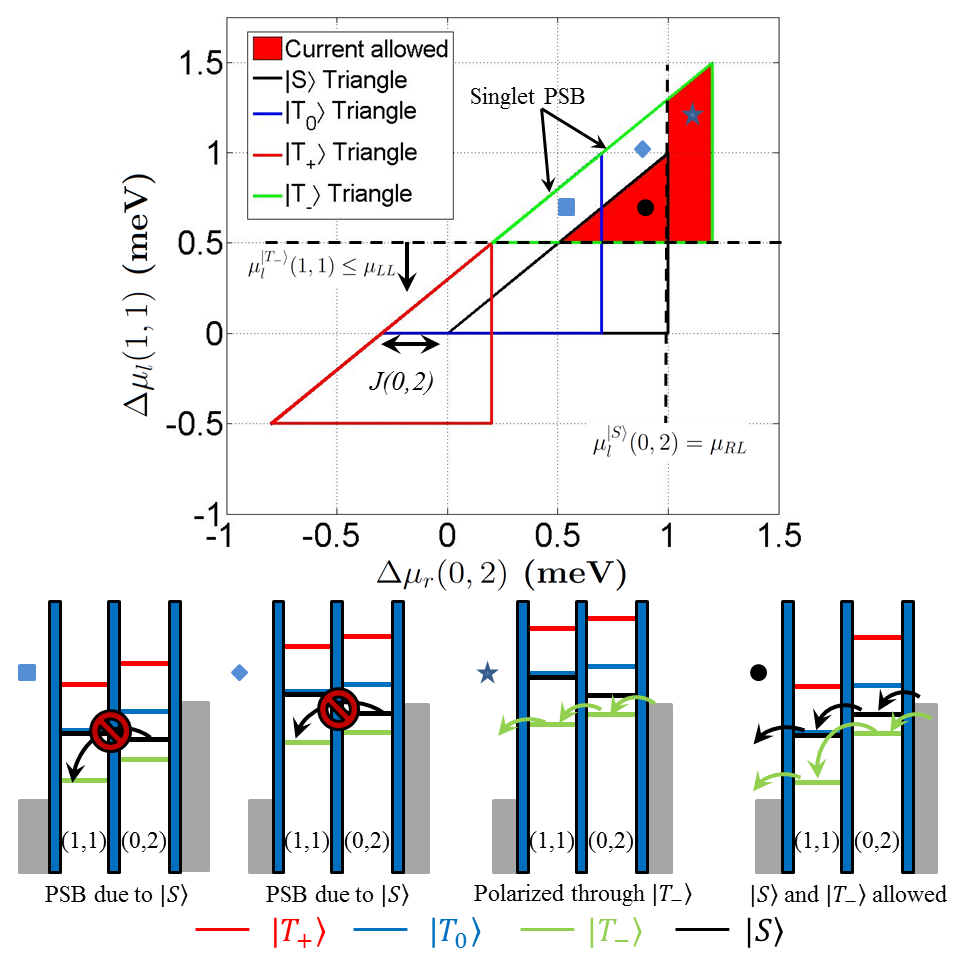}
	\caption{$\mathbf{B \geq B_{ST}}$ \muspace picture of transport in the reverse \reverse direction with an applied magnetic field $g \mu_B B = 0.5~\rm{meV}$, $J(0,2) = 0.3~\rm{meV}$ and $V_{bias} = -1~\rm{mV}$. The region of allowed current (red) is no longer triangular.  The tail-like feature, indicated with the star, consists of current through only the \ket{T_-} levels and therefore is a spin polarized current upon exiting the DQD.  The regions identified with the diamond and square symbols define the area of PSB. Unlike the typical PSB, this high-field PSB is caused by the loading of the \ket{(0,2)S} level rather than the \ket{(1,1)T} level. This means that any leakage current measured in this region is due to relaxing from the \ket{(0,2)S} and \ket{(0,2)T_-} states. In the region identified by the black circle current is allowed through both the  \ket{S} and \ket{T_-} states.  Below the \muspace plot are chemical potential diagrams for the indicated positions with key transitions indicated with arrows.\label{fig3}}
\end{figure}

Both of the high field features can be understood by considering figure~3, a \muspace picture of a transport measurement in the reverse \reverse direction with $E_Z > J(0,2)$. Included are the state triangles for the various singlet and triplet chemical potential levels. The regions labeled with the blue diamond and square correspond to the region of PSB.  The physics of this PSB is similar to the traditional low field \forward PSB, but with the \ket{(0,2)S} state filling the role of the trapped state and the \ket{T_-} levels contributing to current.  This is shown in the corresponding chemical potential diagrams. Throughout the PSB region, although current is allowed through the ground state \ket{T_-} levels, the \ket{(0,2)S} state can load from the right lead and, since the \ket{(0,2)S}$\rightarrow$\ket{(1,1)T_-} tunneling event is not allowed, it will block current until it relaxes to a \ket{T_-} state. Much like how the leakage current in the low field \forward PSB region was used to probe the \ket{(1,1)T}$\rightarrow$\ket{(1,1)S} relaxation\cite{Lai11,Koppens05,Yamahata12,Churchill09A,Churchill09B,Pfund07,Nadj10}, leakage currents in this reverse PSB region can be used to probe the \ket{(0,2)S}$\rightarrow$\ket{(0,2)T_-} relaxation.  

Figure~3 has another distinct feature, the tail region labeled with the star.  This feature also develops when the \ket{T_-} state becomes the ground state in the (0,2) configuration. The feature can be understood by considering the border between the tail and the PSB region. To the left of this border no current is measured because the \ket{(0,2)S} can load and prevent current flow as previously discussed. To the right of the border the \ket{(0,2)S} level is above $\mu_{RL}$ and is therefore unable to load (compare the chemical potential diagrams labelled with the diamond and the star).  Since the \ket{(0,2)T_-} state is the only state that can load, the current measured in the tail region will be spin polarized upon exiting the DQD, which suggests a potential use as a polarized spin current source.\footnote{The current is polarized upon exiting the double quantum dot, however, the spin relaxation process of the electrons once in the Fermi sea of the lead is likely to be fast, quickly diminishing the polarization of the current.}

By thoroughly considering the \muspace picture typically used to explain PSB, we have identified several interesting and unexplored features involved in the \charge transition of a double quantum dot at both low and high magnetic fields.  The boundary between the two field regimes occurs at $B_{ST}$ where the Zeeman energy of \ket{(0,2)T_-} equals $J(0,2)$.  With $g=2$ and exchange energies reported\cite{Lai11,Shaji08,Liu08} in the range of 0.24 to 1.4~meV, $B_{ST}$ in Si should be in the range of 2 to 12~Tesla, accesible with standard cryomagnetics.  This suggests that Si offers an opportunity to test our predictions.  Simple validation can be achieved by performing measurements of the type reported in \cite{Lai11} for both biasing directions and comparing with our predictions of changing triangle size.  Following that, the high-field tail feature we predict should be a clear signature that the reverse PSB regime has been reached.  This naturally leads to experiments involving the leakage current in the PSB regime. Similar to low-field PSB leakage current investigations\cite{Lai11,Koppens05,Yamahata12,Churchill09A,Churchill09B,Pfund07,Nadj10,Johnson05A}, these measurements would probe singlet-triplet relaxation mechanisms but in the (0,2) charge configuration rather than the (1,1). This removes any spin-orbit interaction and nuclear field gradients between the two electrons, both of which were previously shown as the dominant coupling mechanisms between the \ket{S} and \ket{T_x} states in certain material systems\cite{Churchill09A,Nadj10,Pfund07,Yamahata12,Danon09,Nadj10,Pfund07,Churchill09A,Churchill09B,Koppens05}.  Thus, measurements of this type will deepen our understanding of higher order spin relaxation mechanisms.

\begin{acknowledgements}
	We are grateful for the many useful discussions with Josh Pomeroy, Garnett Bryant, Mark Stiles and Michael Gullans.
\end{acknowledgements}

\bibliography{master}

\end{document}


\title{Supplemental material for A new Regime of Pauli-Spin Blockade.}
\author{Justin K. Perron}
\affiliation{California State University San Marcos, Department of Physics, California 92096}
\affiliation{Joint Quantum Institute, University of Maryland}
\affiliation{National Institute of Standards and Technology}
\author{M. D. Stewart, Jr.}
\author{Neil M. Zimmerman}
\affiliation{National Institute of Standards and Technology}
\date{\today}

\maketitle

\renewcommand{\thesection}{\arabic{S}.\arabic{section}}
\section{Allowed Current when $J(0,2) \geq E_{Z}$}

In figure~2 of the main text we show the region of allowed current in a \muspace picture for low magnetic fields. Here, we elaborate on the details of where and why current is allowed. The process of making chemical potential space plots is rather straightforward. For each state, one right-angle triangle is formed with horizontal and vertical sides of length $eV_{bias}$. The singlet triangle has the appropriate vertex positioned at the origin while the other state triangles are shifted according to their appropriate exchange and Zeeman energies. This is shown in figure~S1 with $V_{bias} = -1$~mV, $J(0,2) = 0.3$~meV, and $E_Z = 0.15$~meV.

\renewcommand{\thefigure}{S\arabic{figure}}
\begin{figure}
	\centering
	\includegraphics[width = 0.8\columnwidth]{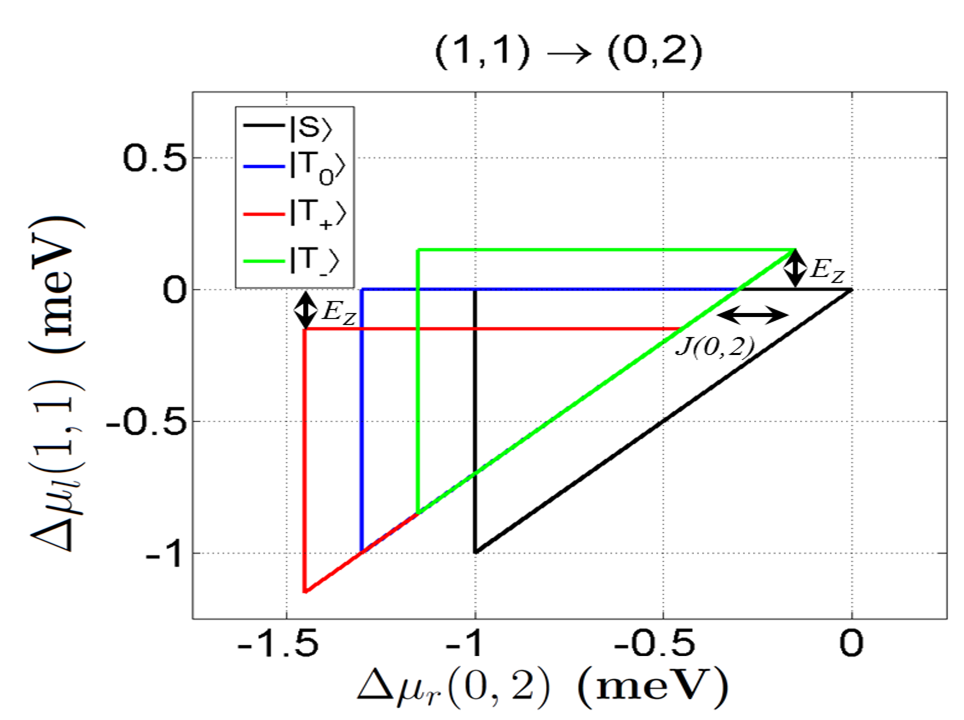}  
	\caption{{\bf{Low field state triangles:}} State triangles in the \forward direction in \muspace with  $J(0,2) = 0.3$~meV, $E_{Z} = 0.15$~meV, and $V_{bias}~=~-1~\rm{mV}$. The singlet triangle always lies at the origin by definition, and all other state triangles are shifted along the two axis according to the appropriate energy splittings.}
\label{figs1}
\end{figure}

To determine the regions of allowed current one can draw the state triangles for the magnetic field of interest and then examine the different regions they define. By counting the conduction paths formed by transitions that are both energetically favorable and allowed through spin conservation one can determine how many channels are allowed in a given region. Secondly, one must determine if there are any partial channels, that is states that can be loaded onto the double dot but not contribute to further tunneling, such as the \ket{(1,1)T_x} levels in traditional PSB. If one of these states exists current is not expected. Figure~S2 shows this process for the \muspace plot in figure~S1. The filled regions are colored according to the number of channels conduction can occur through.  The red region is the PSB region, where, although conduction is allowed through the singlet channel, no current is expected to flow due to the loading of the \ket{(1,1)T_-} state.  In the green region all four states contribute to conduction.  The cyan region corresponds to where the \ket{(1,1)T_+} level cannot be loaded but all other states contribute to current.  Finally, the yellow region corresponds to where there are no blocking partial channels and only the \ket{T_-} levels contribute to conduction. 

\renewcommand{\thefigure}{S\arabic{figure}}
\begin{figure}
	\centering
	\includegraphics[height = \columnwidth]{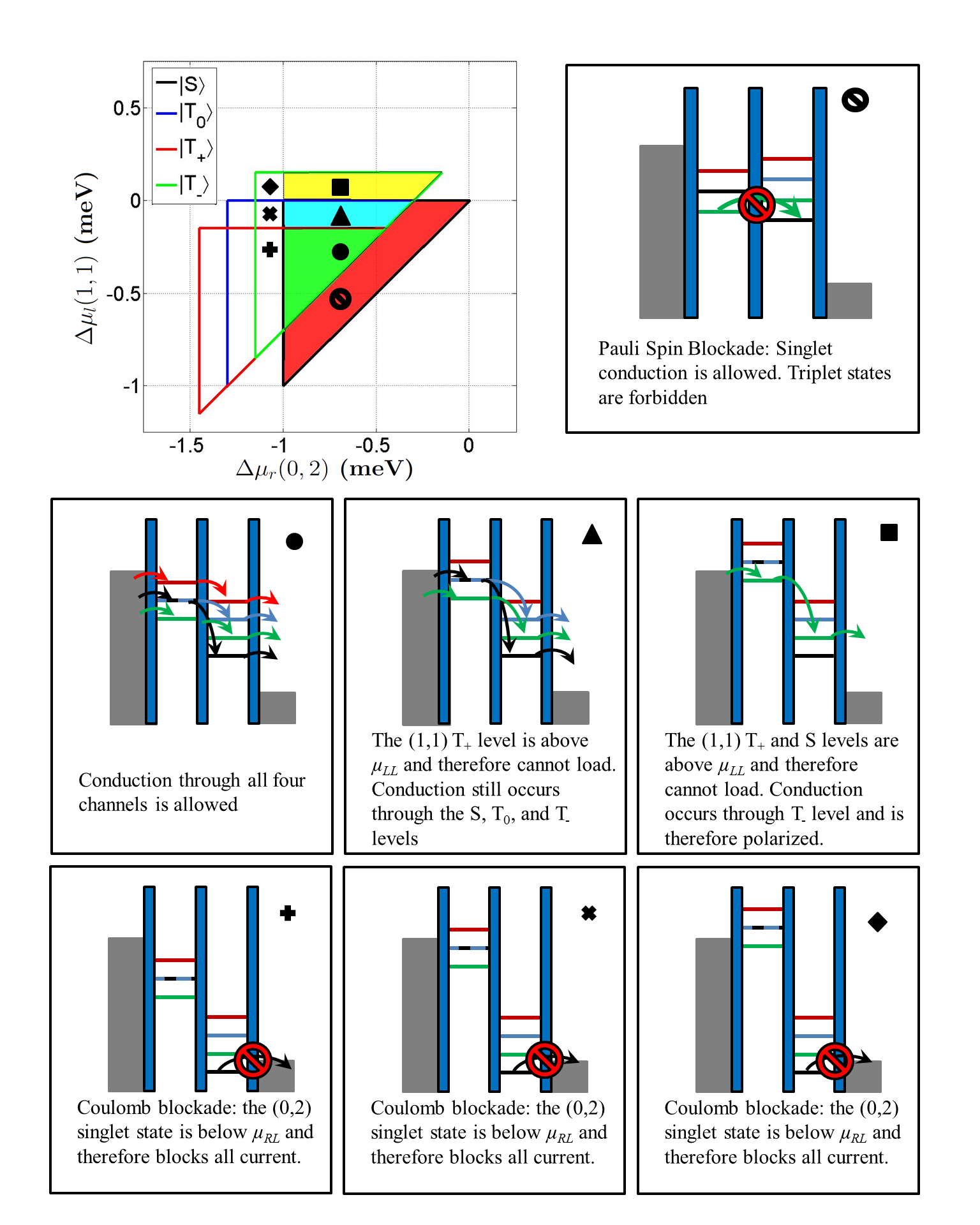}  
	\caption{{\bf{Low field state triangles:}} State triangles in the \forward direction in \muspace with  $J(0,2) = 0.3$~meV, $E_{Z} = 0.15$~meV, and $V_{bias}~=~-1~\rm{mV}$. Colors in the filled region correspond the number of allowed conduction channels. The red region is the Pauli-spin blockade region with one allowed channel but one blocking partial channel. In the green region conduction is allowed through all four states, \ket{S}, \ket{T_0}, \ket{T_+} and \ket{T_-}. In the cyan region the \ket{T_+} level cannot load so it is not involved in conduction whereas in the yellow region only the \ket{T_-} level can load. Thus, current is polarized. Also shown are chemical potential diagrams corresponding to various points in the plot, which indicate whether or not current is allowed.}
\label{figs2}
\end{figure}

\section{Internal Structure}
Looking at figure S1, one might expect some current dependence relating to the number of conduction channels allowed.  Any dependence would likely depend heavily on any state dependent tunnel rates and very device specific.